\documentclass[a4, reprint,showpacs,showkeys,prb,aps,noeqsecnum]{revtex4-1}
\usepackage{graphicx}
\usepackage{subfig} 
\usepackage{epstopdf}
\usepackage{bm}
\usepackage{amsmath}
\usepackage{mathrsfs}

\begin{document}

\title{
A Monte Carlo study of critical properties\\ of strongly diluted magnetic semiconductor (Ga,Mn)As
}

\author
{P. Tomczak, H. T. Diep$^\dag$, P. Jab{\l}o{\'n}ski, H. Puszkarski}
%\email{ email: }
\affiliation
{Faculty of Physics, ul.~Umultowska 85, \\
Adam Mickiewicz University, 61-614 Pozna\'n, Poland\\
$^\dag$Laboratoire de Physique Theorique et Modelisation\\
Universite de Cergy-Pontoise, CNRS, UMR 8089 2,\\
Avenue Adolphe Chauvin, 95302 Cergy-Pontoise Cedex, France
}

\date{\today}

\begin{abstract}
Within a Monte Carlo technique we examine critical properties of 
diluted bulk magnetic semiconductor (Ga,Mn)As 
modeled by a strongly diluted ferromagnetic Heisenberg spin-$\frac{5}{2}$
system on a face centered cubic lattice. We assumed that 5\%
of Ga atoms is substituted by Mn atoms and 
the interaction between them is of the RKKY-type.
The considered system is randomly quenched and a double average was performed:
firstly, over the Boltzmann probability distribution and secondly - over
2048 configurations related to the quenched disorder.
We estimated the critical temperature: $T_c=97\pm6$ K, which is in agreement with the experiment.
The calculated high value of critical exponent $\nu$ seems to point to 
a possibility of non-universal critical behavior.
\end{abstract}

%\keywords{ferromagnetic semiconductors, (Ga,Mn)As thin films, spin-wave resonance, 
%surface anisotropy, surface spin pinning, surface exchange length}

\maketitle

% ---------------------------------------------------------------------------

%{\em Introduction.} --- 

Gallium manganese arsenide, (Ga,Mn)As, is a 
diluted magnetic semiconductor with a zinc-blende crystal structure with two
interpenetrating FCC lattices (Ga and As). In one of them several percent of Ga
is substituted, probably randomly, by Mn. This leads to the strong {\em quenched} site disorder.
The compound is still focusing a lot of theoretical and experimental
attention mainly due to its potential spintronic applications (i.e., a possible manipulation 
the spin and the charge carrier degrees of freedom at the same time). 
The fundamental question while studying collective spin 
phenomena
is the question about the critical properties i.e., critical temperature and possibly - which is more difficult -
about critical exponents.
The effects of quenched disorder on critical properties
of such systems have been the subject of intense 
experimental and theoretical interest for a long time.
Generally, the critical temperature of bulk (Ga,Mn)As
depends on the concentration of Mn atoms, but details 
of this dependence are 
far from being known. Let us point to the recent\cite{Wang} 
experimental answers, e.g., for the Mn concentration of 12\% 
the value of $T_C$=183.5 K has been  obtain for  samples from remanent magnetization
Kouvel-Fisher plots.
On the other hand,
using massive Monte Carlo simulations the critical properties of strongly disordered (but above
the percolation threshold) Heisenberg systems [being model of (GaMn)As] on  
{\em simple cubic lattice} were examined\cite{Sarma}. Critical temperatures and critical exponents 
were estimated and it was shown, that Harris criterion is fulfilled.

In this paper we would like to elucidate the critical properties of 
strongly (the concentration of Mn atoms is below the percolation threshold) 
diluted (Ga,Mn)As.
These systems
are usually examined theoretically using a classical Heisenberg model,
\begin{equation}
H = -\sum_{\substack{i,j}} J(r_{ij}) {\bf S}_i {\bf S}_{j}.
\label{Hamiltonian}
\end{equation}
$J(r_{ij})$ stands for the hole-mediated, indirect exchange coupling 
between Mn moments separated by a distance $r_{ij}$ on FCC lattice,
\begin{equation}
J(r) =  J_0\mbox{e}^{-r/l}r^{-4}\Big(\mbox{sin}(2k_Fr) - 2k_Fr\,\mbox{cos}(2k_Fr)\Big).
\label{RKKY_Int}
\end{equation} 
$k_F$ is the Fermi wave number $k_F=(\frac{3}{2}\pi^2\,n_c)^{1/3}$, $n_c$ stands for the hole density
and $l$ is the damping scale.

\begin{figure}[!h]
\centering
   \includegraphics[width=0.25\textwidth, angle=0]{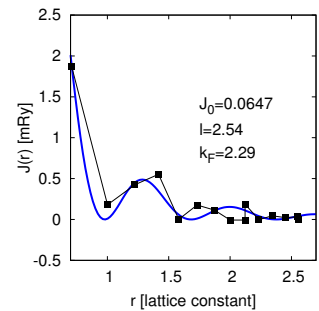}
   \caption{Exchange interaction $J(r)$ in (Ga,Mn)As with 5\% of Mn concentration versus distance, 
   as calculated by Sato\cite{Sato} (black squares). The blue line is a result 
   of fitting Sato's data to Eq.~\ref{RKKY_Int} 
}
   \label{RKKY_dist}
\end{figure}

Many authors have estimated exchange interactions entering to Eq.~\ref{RKKY_Int}.
Sato\cite{Sato}, using KKR method calculate (Ga,Mn)As electronic structure
and subsequently embedding Mn impurities in CPA medium was able to
to find the $J(r)$, see Fig.~\ref{RKKY_dist} (black squares).
Fitting his data to the phenomenological formula given by Eq.~\ref{RKKY_Int} one can estimate 
the period of RKKY oscillations, $\sim\pi/k_F\approx1.37$ (cf. discussion in Ref. [4]).

To obtain critical properties we have used an approach based on
finite-size scaling hypothesis and 
analyzed the scaling of two quantities,
which do not depend on the scale at which we look at the system: 
reduced correlation length $\frac{\xi}{L}$ and Binder cumulant $U_2$.
As a test of this approach we have estimated the critical temperature and critical
exponents $\nu$ and $\beta$ for classical FCC Heisenberg ferromagnet nearest neighbor
interactions only.
Firstly, let us recall the definition of the correlation length $\xi$ 
using the structure factors for two wave vectors: ${\bf q}=[0,0,0]$ and ${\bf q_1}=[\frac{2\pi}{L},0,0]$
\begin{equation}
\xi = \frac{1}{|{\bf q_1}|}\sqrt{\frac{S({\bf q})}{S({\bf q_1})}-1},
\label{ksi}
\end{equation}
and the structure factor is given by
\begin{equation}
S({\bf q}) = \sum_{\bf r} \mbox{cos}({\bf q}\cdot{\bf r}) C({\bf r}),
\label{sq}
\end{equation}
with $C({\bf r})$ being the correction function at the distance $|{\bf r}|$.
Subsequently, using the Metropolis algorithm (200000 MC steps/spin, 10000 steps/spin to reach
equilibrium) we have calculated  $\xi(T)$ for three FCC systems:
$L$=10,\,12,\,14 consisting of 4000,\,6912 and 10976 spins, respectively, with periodic
boundary conditions.
\begin{figure}[!ht]
   %\captionsetup[subfigure]{labelformat=empty}
   \centering
   \subfloat{\includegraphics[width=0.23\textwidth]{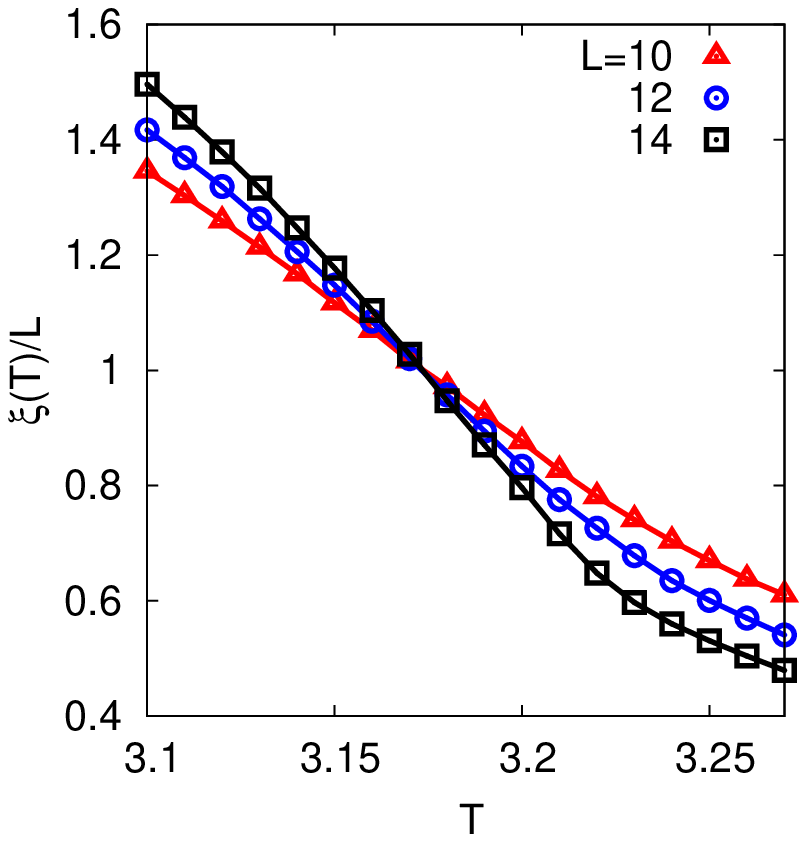}}\quad
   \subfloat{\includegraphics[width=0.23\textwidth]{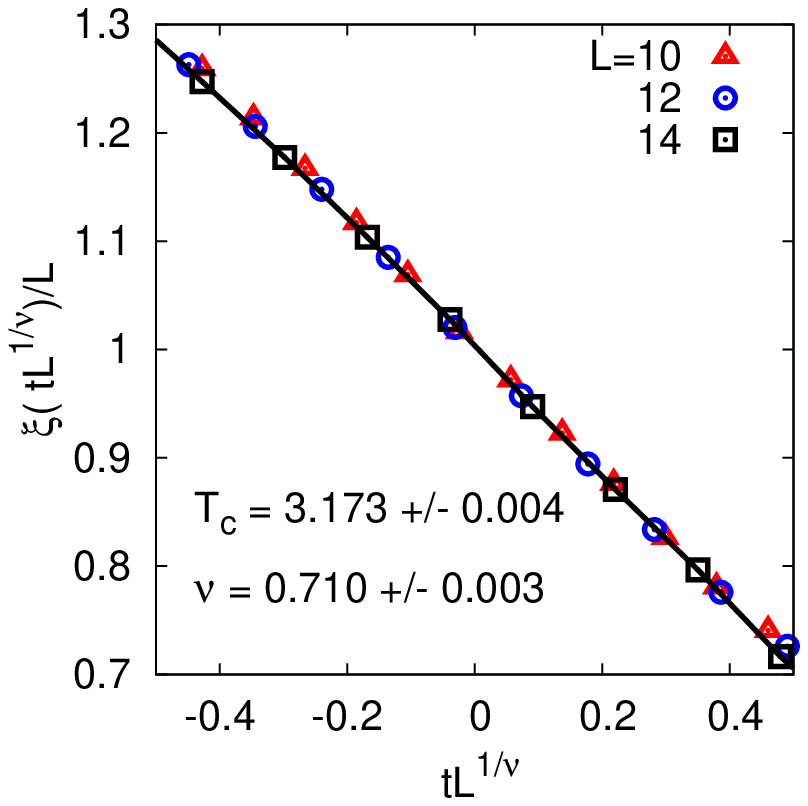}}
   \caption{Temperature dependence of the three correlation lengths 
normalized by the system size L in three classical Heisenberg FCC systems  before rescaling(left)
and after rescaling (right).
$t$ stands for the reduced temperature, $\frac{T_c-T}{T}$.
The scaling collapse leads to the optimal values of $T_c$ and exponent $\nu$.
Statistical errors are comparable with points sizes.}
   \label{ksi_L}
\end{figure}
Since one assumes that a quantity which is singular
at $T_c$ in the thermodynamic limit, scales with the system size $L$ close
to $T_c$ as a power of
$L$ multiplied by a non-singular function of the ratio $\frac{\xi}{L}$, one has
\begin{equation}
\xi = L g(t  L^{\frac{1}{\nu}}),
\label{fss_ksi}
\end{equation}
where $g$ stands for a scaling function. Thus,
an attempt to plot $\xi$ vs $t  L^{\frac{1}{\nu}}$
with a proper values of $T_c$ and $\nu$ should result in a scaling collapse -
this is shown in Fig. \ref{ksi_L} (right).

\begin{figure}[!ht]
   %\captionsetup[subfigure]{labelformat=empty}
   \centering
   \subfloat{\includegraphics[width=0.23\textwidth]{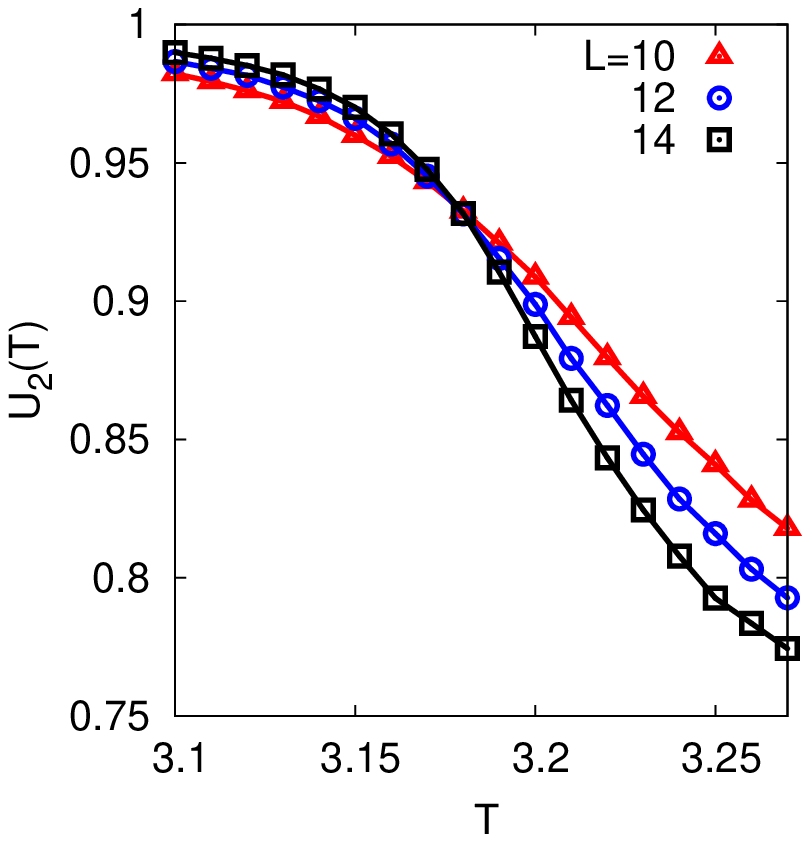}}\quad
   \subfloat{\includegraphics[width=0.23\textwidth]{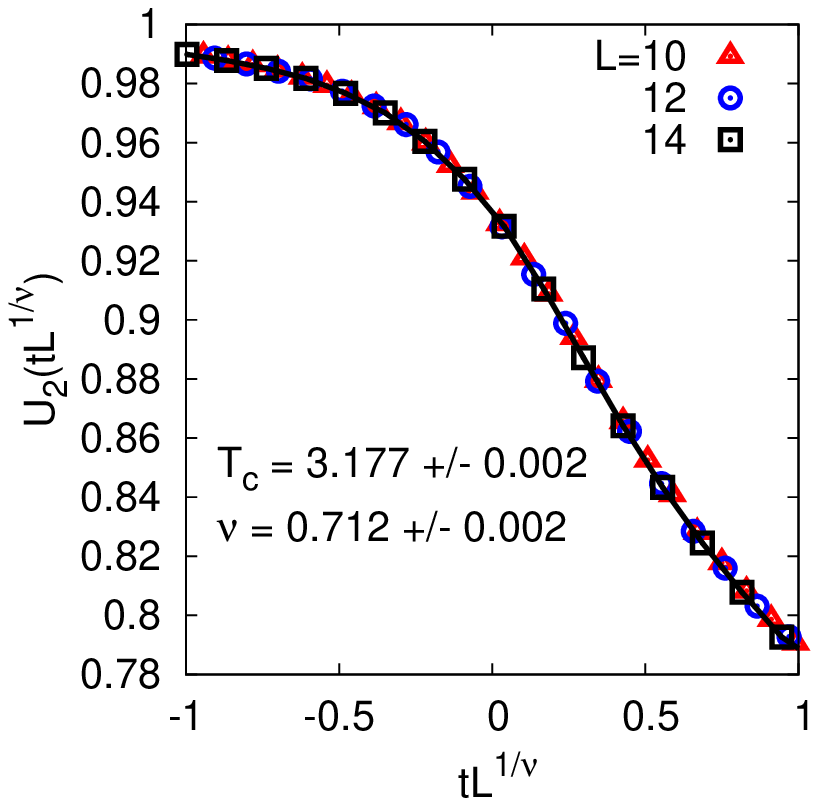}}
   \caption{The cumulant $U_2$ before (left) and after rescaling (right). As expected from the finite size
scaling ansatz (Eq. \ref{U_2_scaling}) the data for different system sizes 
collapse on a single curve for $T_c=3.177\pm0.002$ and $\nu=0.712\pm0.002$ (right).}
   \label{U2}
\end{figure}

Besides examining the scaling of $\frac{\xi}{L}$ there exists other dimensionless 
size-independent quantity at the critical point which can be used
to extract critical properties of the system under consideration, namely the Binder $U_2$ cumulant.
In the case of classical Heisenberg system (or a vector order parameter) it is 
defined in the following way:
\begin{equation}
U_2 = \frac{5}{2}\Big(1 - \frac{3}{5}\frac{\langle m^2\rangle}{\langle |m|\rangle^2}\Big).
\label{U_2}
\end{equation}
During the same MC run we have calculated $\langle m^2\rangle$ and $\langle |m|\rangle$.
 The scaling of $U_2$ cumulant
enables to extract a critical exponent $\nu$ for the second time
from the scaling of the function $h$
\begin{equation}
U_2 = h(t  L^{\frac{1}{\nu}}).
\label{U_2_scaling}
\end{equation}
This is shown on the left side of Fig. \ref{U2} - before rescaling and on the right side - after rescaling. 

\begin{figure}[!ht]
   %\captionsetup[subfigure]{labelformat=empty}
   \centering
   \subfloat{\includegraphics[width=0.23\textwidth]{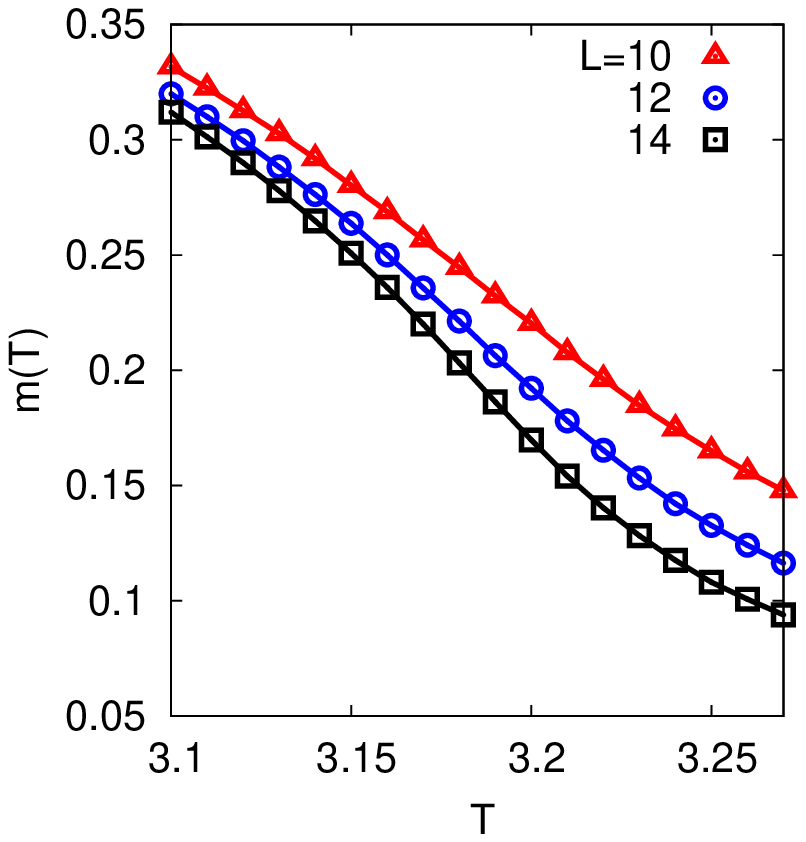}}\quad
   \subfloat{\includegraphics[width=0.23\textwidth]{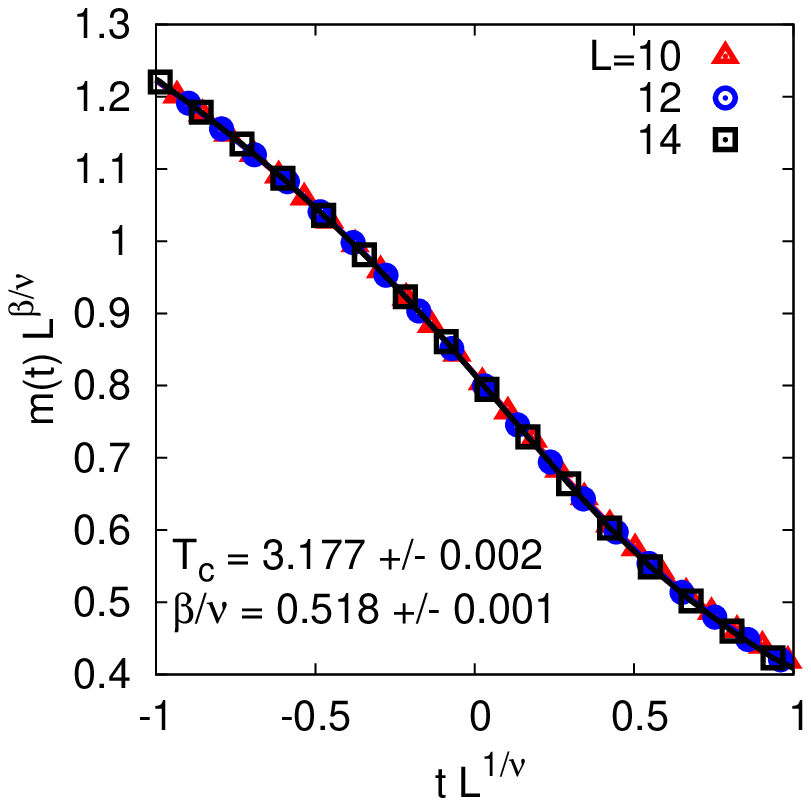}}
   \caption{The magnetization m(T) before (left) and after rescaling (right). As expected from the finite size
scaling ansatz (Eq. \ref{M_scal}), the data for different system sizes 
fall onto a single curve for $T_c=3.177\pm0.002$ and $\frac{\beta}{\nu}=0.518\pm0.001$ (right).}
   \label{m_scaling}
\end{figure}

It is also possible to estimate the second independent exponent $\beta$ from the following %{\em L-dependent}
scaling
of the magnetization
\begin{equation}
m(t,L) = L^{-\frac{\beta}{\nu}} \,f(t L^{\frac{1}{\nu}}),
\label{M_scal}
\end{equation}
with $f$ being some scaling function. Plotting $m(t,L) L^{\frac{\beta}{\nu}}$ 
versus $tL^{\frac{1}{\nu}}$ for different system sizes
leads to the scaling collapse for $\frac{\beta}{\nu}=0.518\pm0.001$, see Fig. \ref{m_scaling}.

Let us summarize our tests concerning critical properties of the classical FCC Heisenberg ferromagnet:
we were able to reproduce the values of critical temperature $T_c=3.177\pm0.002$
and critical exponents $\nu=0.712\pm0.002$, $\beta=0.369\pm0.002$ for relatively small systems with acceptable accuracy 
which is sufficient to conclude whether the Harris criterion is applicable.
Our numerical values should be compared\cite{Pelissetto, Adler} 
with $T_c=3.1771\pm0.0001$, $\nu=0.7112\pm0.0005$ and $\beta=0.3689\pm0.0003$.

Let us now move to a more complicated structure, namely to {\em a strongly diluted} classical
Heisenberg FCC system with {\em long-range RKKY-type} interactions, given by Eq. \ref{RKKY_Int}.
Suppose that a concentration of magnetic Mn atoms is equal to 0.05 which is far below the
percolation threshold $p_0$ for the nearest neighbor Heisenberg
system ($p_0$=0.20 for the FCC lattice).

\begin{figure}[!h]
\centering
   \includegraphics[width=0.25\textwidth, angle=0]{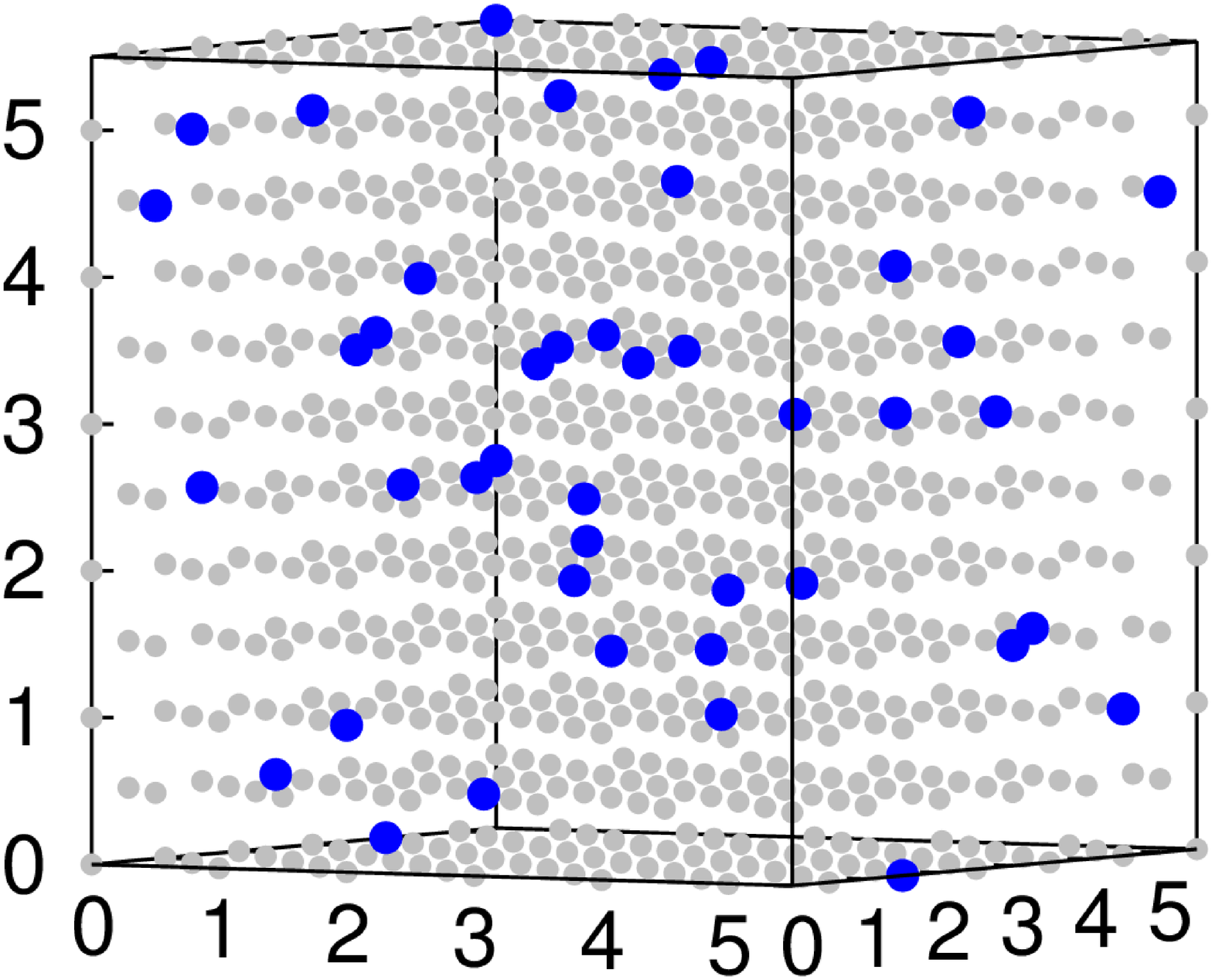}
   \caption{One of the possible configurations of magnetic atoms chosen from
$6^3\cdot4$ atoms (grey) in FCC structure. Only 5\% of them are magnetic ones (blue).
Averaging over such configurations of magnetic atoms one should take into account $\sim\!\!10^4$ of them. }
   \label{GaMnAs_dist}
\end{figure}

The system is a randomly quenched one and therefore
one has to take a double average: for 
fixed configuration of magnetic atoms, see Fig.~\ref{GaMnAs_dist}, one computes the thermal average
within the Metropolis importance sampling
and subsequently, the average over disorder is realized by simple sampling.

One could ask a basic question {\em how large} should be the sample which will be used for averaging
over configurations. Samples being examined in typical experiments amount a very large number 
of degrees of freedom ($\sim\!\!10^{23}$)
and the observable quantities are {\em self-averaging}\cite{Binder}.
the situation is quite different in the case of a finite
size systems and finite size scaling analysis.
The main problem is that by considering systems of finite linear dimension
$L$ at the critical temperature $T_c$
one encounters fluctuations by passing from one
configuration to another which cause a significant fluctuation
of the pseudo-critical temperature $T_c$.
This implies that one has to average over $\sim\!\!10^4$ configurations
in order to get the relative error of the disorder average at
$T_c$  less than 1\%\cite{Binder}. 

\begin{figure}[!ht]
   %\captionsetup[subfigure]{labelformat=empty}
   \centering
   \subfloat{\includegraphics[width=0.23\textwidth]{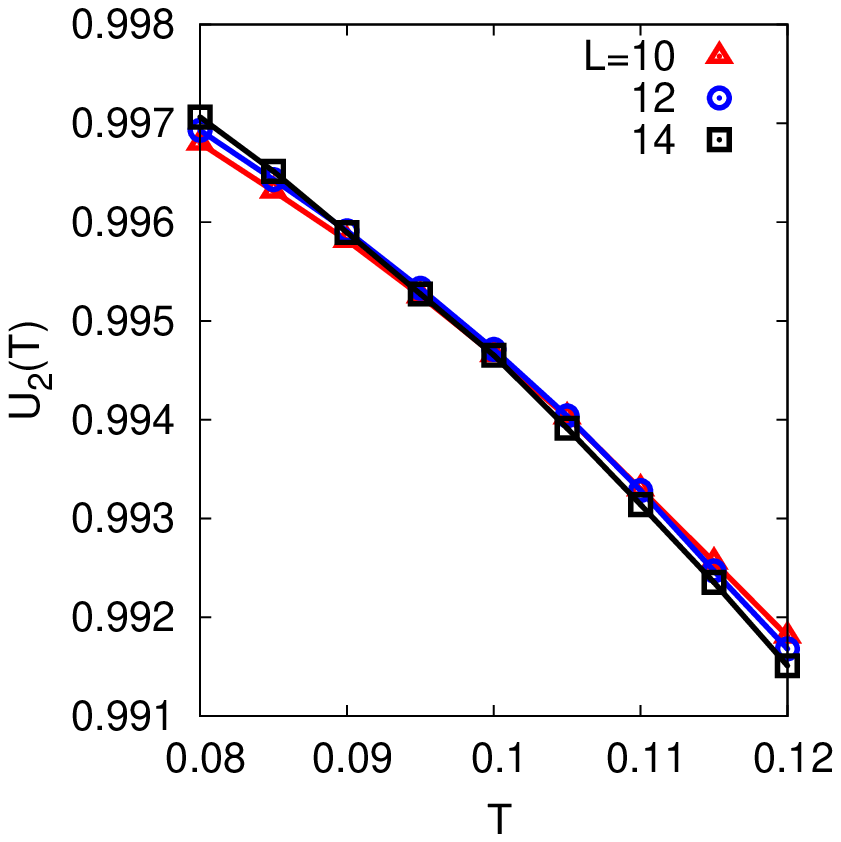}}\quad
   \subfloat{\includegraphics[width=0.23\textwidth]{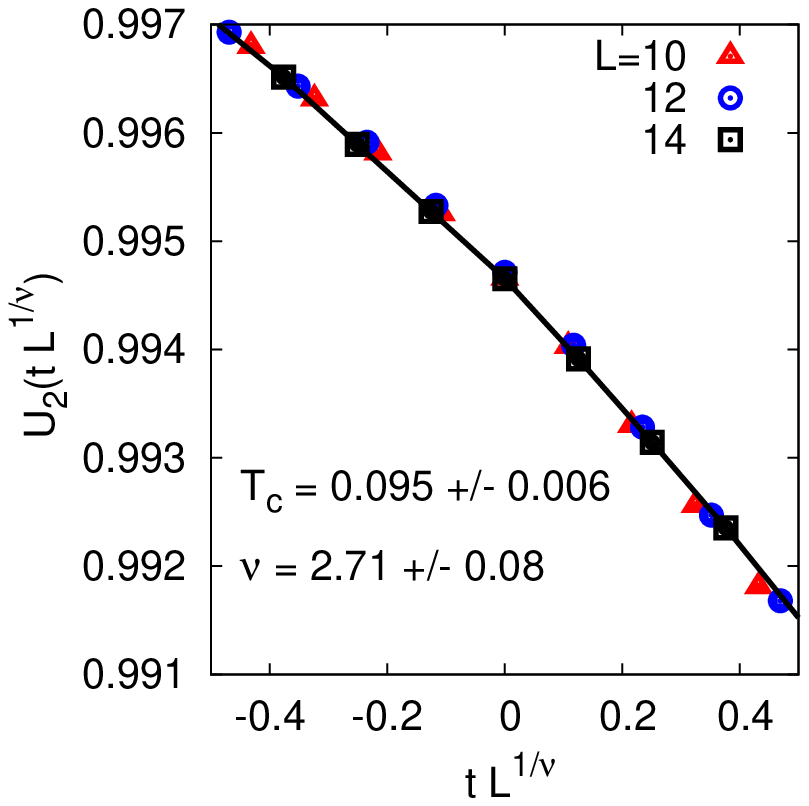}}
   \caption{The magnetization m(T) for diluted systems before (left) and after rescaling (right). 
   The data for different system sizes 
fall onto a single curve for $T_c=0.095\pm0.006$ and $\frac{\beta}{\nu}=0.370\pm0.050$ (right).
The concentration of Mn atoms amounts 5\%, interactions\cite{Sato} are given in Fig.~\ref{RKKY_dist}, black squares.}
   \label{diluted_scaling}
\end{figure}

Let us summarize the examination of the critical properties of (Ga,Mn)As with 5\% Mn atoms and interactions 
between them presented in Fig.~\ref{RKKY_dist}.
The average values of $m(T)$ and $U_2(T)$ were calculated for 2048 configurations. It was not possible 
to find a proper scaling dependencies $m(T)$ -
statistical errors resulting from fluctuations  were much bigger than differences between $m_{10}(T)$, $m_{12}(T)$ and  $m_{14}(T)$.
However, it was possible to find a proper scaling for the Binder cumulant $U_2(T)$. The effect (crossing in Fig.~\ref{diluted_scaling} - left) is rather subtle but clearly marked
and it is definitely beyond the statistical error. 
Taking into account that 1 Ry = 158 K and $S=\frac{5}{2 }$ for Mn, one can estimate $T_c$ for (Ga,Mn)As containing of 5\% Mn atoms
as follows:
$T_c=158 \cdot \Big(\frac{5}{2}\Big)^2 \cdot 0.095 = 97\pm6\,\,$K which is in agreement with experiment\cite{Wang}. 
Let us stress that both types of data, presented in Fig.~\ref{RKKY_dist} lead to the same (within a statistical error) value of $T_c$.
A high a value of the exponent $\nu=2.71$   may indicate
that the considered transition belongs to a universality class different from that of the 3D Heisenberg.
Possible long-range RKKY interactions and strong site quenched dilution might be the cause of
this difference. Similar behavior was observed\cite{Kawamura} in case Heisenberg spin glasses.
We leave this issue open for future investigation.

{\em Acknowledgments.}\!\! --- 
This work was a~part of a~project financed by Narodowe
Centrum Nauki (National Science Centre of Poland),
Grant No. DEC-2013/08/M/ST3/00967. Numerical calculations were performed 
at Pozna\'n Supercomputing and Networking Center under Grant No. 284.

% ------------------------------------------------------------------
% Figures

\newpage

\end{document}